\documentclass[11pt,a4paper]{article}
\usepackage{epsfig,amsmath,amssymb,array,dcolumn,subfigure,rotating}
\def\ul#1#2{\textstyle{\frac#1#2}}
\begin{document}

\title{Two-body polyelectrolyte mediated  bridging interactions}
\author{Rudi Podgornik \\[3mm] 
Laboratoire de Physique des Solides \\
Universite Paris Sud, Bat 510 \\
91405 Orsay Cedex, France\\
and\\
Department of Physics, University of Ljubljana,\\
Jadranska 19, 1000 Ljubljana, Slovenia
}
\maketitle

\begin{abstract}
We investigate theoretically polyelectrolyte bridging interactions on the two-body level. The model system is composed of two macroions with two oppositely charged flexible chains. The electrostatic interactions are treated on the Debye - H\" uckel level. The formal level of  the theory is provided by the Feynman-Kleinert variational method generalized to include also self-interactions between polyelectrolyte segments. The variational equations are shown to exhibit two solution branches corresponding to {\sl strong} and {\sl weak} coupling,  whereas  conformations of the chain can be described as {\sl weakly} or {\sl strongly paired}. We investigate the effective pair interaction between the macroions in the parameter space and comment on the relevance of the calculation for bridging interactions in experimental context. 
\end{abstract}

\section{Introduction}

Polyelectrolytes are ubiquitous in colloidal systems and play a fundamental role in determining the interactions between as well as stability and structure of various molecular assemblies \cite{generalpoly}. Their effect on colloidal interactions has been studied and exploited in various technological contexts ranging from paper industry to pharmaceutical industry \cite{applic}. It seems however that their most basic role is played in the biological context where their importance can be hardly overestimated. They are an essential and fundamental component of the cellular environment and make their mark in its every structural and functional aspect \cite{bio}. 

The behavior of polyelectrolytes in biological context has without any doubt been one of the foci of the soft matter research for quite a few years now \cite{leshouches}. The intense work on the interactions and mesophase behavior of the most studied polyelectrolyte in the biological context if not in general, {\sl i.e.} DNA, has elucidated many fascinating physical aspects of this molecule and the repercussions that they have on the structure and function of biological matter \cite{currop}. The mesoscopic interactions between many DNA molecules \cite{currop1} and elastic properties of single DNA molecules in different solution conditions  \cite{rouzina} have been measured directly and are understood on a fundamental physical level. Not all biological polyelectrolytes or all solution conditions have been or indeed can be studied at quite the same level of detail. Sometimes a lot less information then direct measurement of molecular interactions at all macromolecular densities is experimentally available. Studies at low macromolecular densities in systems where polyelectrolyte behavior is expected to show its mark usually only lead to second virial coefficients and not complete interaction curves \cite{wissenburg, petsev, raspaud}. This is not due to poorly designed experimental setup but shows a rather fundamental limit in the amount of information that can be provided by experiments at these conditions.  In this situation one has to rely heavily on different models of the mesoscopic interaction potential and the way they transpire through the measured second virial coefficient. This is the first point of departure of this work.

The second one is a very peculiar interaction in polyelectrolye systems, where long charged polymers can mediate interactions between macroions of opposite charge \cite{andelman1} (and references therein). The term {\sl bridging interactions} is usually applied to this situation where a single chain can adsorb to different macroions and via its connectivity and elasticity mediate attractive interactions between them. These interactions have been studied intensively both experimentally \cite{experiments} as well as theoretically  \cite{podgornik1, theory, podgornik2, borukhov}. Surface force apparatus and atomic force microscopy have provided direct data on the separation dependence of the bridging interaction between macroscopic surfaces with polyelectrolyte chains either grafted or in chemical equilibrium with a bulk solution  \cite{experiments}. Theoretical work has added a clear mesoscopic picture for the bridging interaction between macroscopic surfaces and elucidated the effects of salt and non-electrostaic excluded volume effects on the strength and range of this interaction \cite{podgornik1, theory, podgornik2, borukhov}. Since it is based on sometimes severe model or formal restrictions there is no single theoretical approach that is able to account for all experimentally observed details or is able to explore in comparable details all the regions of the parameter phase space \cite{andelman1}. The fact that the effects of bridging interaction between small macroions \cite{podgornik2} as opposed to macroscopic surfaces \cite{andelman1} have been studied to a much smaller extent thus makes a strong point for its reevaluation. 

The main motivation for this task are recent experiments on the second virial coefficient of the nucleosomal core particles (NCP) in low density regime at various solution conditions \cite{raspaud, thesis}. NCPs represent the lowest level of the chromatin organization in eucaryotes and have recently been resolved at an atomic resolution \cite{luger}. They consist of a histone protein octamer core with 146 bp of DNA tightly wrapped around it, giving it an approximate cylindrical shape of a radius $\sim$ 55 \AA, a height of $\sim$ 57 \AA~ and a structural charge of $\sim$ -250. This complex is stable in an aqueous solution from mM to 750 mM monovalent salt ionic strength \cite{thesis}. The  charged histone N-termini or N-tails can desorb from this complex and basiaclly play the role of  grafted flexible polyelectrolyte chains of an approximate total charge of $\sim$ +90.They remain essentially adsorbed to the DNA segment of the NCP at low ionic strength but  tend to assume a more extended conformation as the ionic strength is increased \cite{raspaud}. The application of classical \cite{parsegian} and manometric \cite{raspaud1} osmometry provided data on the osmotic pressure of the NCP in NaCl solutions of variable ionic strength from which  the second virial coefficient was deduced quite accurately \cite{raspaud}. It was demonstrated that the second virial coefficient is a non-monotonic function of the ionic strength and a plausible though tentative hypothesis was given, that the non-monotonicity of the second virial coefficient might be due to the bridging interaction mediated by the extended N-tails of the NCPs \cite{raspaud}. Very similar non-monotonic dependence of the second virial coefficient was seen also in apoferritin solutions where a bridging interpretation is hard to envision \cite{petsev}. 

Motivated by this important experimental result and its tentative interpretation we embarked on a detailed study of the interaction between charged macroions with grafted oppositely charged polyelectrolyte tails, as a function of the ionic strength of a monovalent bathing salt solution. The level of the theoretical calculation had to be considered,  that would allow for a straightforward evaluation of the second virial coefficient, being an integral of the underlying interaction potential,  and its dependence on the ionic strength. Also since the N-tails in the motivating experiment can not be considered as infinitely long, {\sl i.e.} there are finite size effects that need to be taken properly into account, it seems that a mean-field theory of bridging interactions as formulated for the case of interacting charged planes and based on the ground-state dominance {\sl ansatz}  \cite{podgornik1, borukhov} can not be simply implemented to the present case. Finite size effects are quite difficult to deal with on the MF level, especially if one needs to evaluate the interaction potential between macroions in a very large (ideally infinite) range of separations. In view of all this we formulated a variational \cite{podgornik3} , two particle (two macroions) theory of the bridging interaction that starts from an explicit mesoscopic ployelectrolyte model and includes the interactions of the polyelectrolyte chain with the macroions, the interactions of the chain with itself and connected with it the effect of the electrostatic stiffening of the chain, as well as the configurational entropy of the chain. In some respects the theory proposed here could be viewed upon as a variational version of the Asakura - Oosawa theory \cite{asakura}. The finite size of the chain can be relatively straightforwardly dealt with on this level of the theory and gives rise to important features of the two-particle bridging interaction that are lost in the simplest, ground state dominance formulation of the MF theories.

The organization of the paper is a s follows: we will first describe the model and give an introduction to a modified variational Feynman-Kleinert approach to the polyelectrolyte chains. We will derive the main equations and solve them numerically for different conditions. We will show that in general the bridging interactions for this model system comes in two varieties that we dub the {\sl strong} and the {\sl weak} coupling limit. The form of the total interaction between the macroions will be obtained numerically for all values of the intermacroion separation as a function of system parameters such as the amount of fixed charge on the macroions, the length of the polyelectrolyte chain and the screening length of the intervening bathing solution. We will discuss the ramifications of these results for the salt dependence of the second virial coefficient and point to the possible shortcomings of the calculation and guess about the way to possibly circumvent them. We will finally comment on the significance of the present calculation for the understanding of the bridging interaction in the NCP solution system.

\section{Model}

The model system that we take as the starting point of our theoretical investigation of the bridging interaction is quite simple: it is composed of two spherical point macroions with M negative fixed charges plus two oppositely charged chains, each with $N$ monomers, one per each monomer.  
The pair interaction potential $u({\bf r}', {\bf r})$ between all the charges in the system will be taken of the screened Coulomb (Debye-H\" uckel) form \cite{generalpoly}
\begin{equation}
	u({\bf r}', {\bf r}) = \frac{e_0^2}{4\pi \epsilon\epsilon_0} \frac{e^{-\kappa \vert {\bf r} - {\bf 	r}'\vert}}{\vert {\bf r} - {\bf r}'\vert} \qquad {\rm or ~else} \qquad u({\bf k}) = \frac{e e'}{\epsilon\epsilon_0 (k^2 + \kappa^2)},
	\label{eq.0.1}
\end{equation}
in real in in the Fourier space, the form we will need later on. $\kappa$ is the inverse Debye length, $e_0$ is the elementary charge, one per each Kuhn's length and the rest of the notation is standard. Obviously counterions are not explicitly included in this model. The interaction potential between the polyelectrolyte and the macroion charges is assumed of a similar form {\sl viz.} 
\begin{equation}
	\phi_{ext}( {\bf r}) = \frac{e_1 e_0}{4\pi \epsilon\epsilon_0}  \frac{e^{-\kappa \vert {\bf r} - {\bf 	r}_1\vert}}{\vert {\bf r} - {\bf r}_1\vert} +  \frac{e_2 e_0}{4\pi \epsilon\epsilon_0}  \frac{e^{-	\kappa \vert {\bf r} - {\bf 	r}_2\vert}}{\vert {\bf r} - {\bf r}_2\vert} + \dots, 
	\label{eq.0.2}
\end{equation}
where ${\bf r}_1$, ${\bf r}_2$ etc. are the positions of macroions and their charges are $e_1 = e_2 = M e_0$ etc.. Our model is thus a very straightforward generalization to many macroions of the model used in polyelectrolyte adsorption studies \cite{stoll}.

\begin{figure}[ht]
\begin{center}
    \epsfig{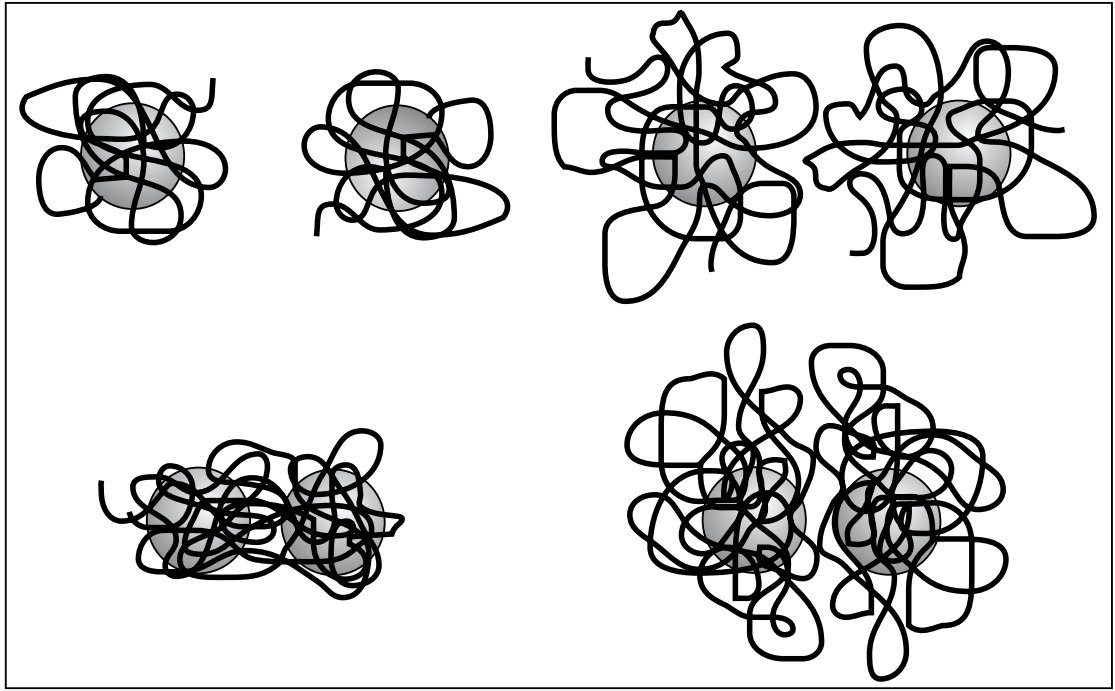}
\end{center}    
\caption{A schematic representation of the model and the nature of the variational solutions. L.h.s.: the {\sl strong coupling} solution branch. The conformation of the chain is determined mostly by the interaction with the two macroions. It can be either in the {\sl weakly} (upper) or {\sl strongly paired} (lower drawing) configuration, depending on the separations between the macroions. The attractive bridging interaction is strongest in the {\sl strongly paired} configuration. R.h.s.: the {\sl weak coupling} solution branch. The conformation of the chain is determined mostly by the self-interactions of the chains. They are in the  {\sl weakly paired}  configuration for any separation between the macroions. On approach the interpenetration of the monomer clouds (the "coronas") leads to prevalently repulsive interactions. Note that the macroions are treated as point aprticles in the claculation and that their size is thus exagerated in this drawing.}
\label{fig0}
\end{figure}

We will use a standard (Edwards) model \cite{cloizeaux} for the polyelectrolyte chain where the mesoscopic Hamiltonian has contributions from chain connectivity, interactions between the segments of the chain and the interaction with an external field due to the presence of two macroions. It is written as
\begin{equation}
	\beta {\cal H}[{\bf r}_i(n)] = \frac{3}{2 \ell^2} \sum_{i=1}^2 \int_0^N \dot{\bf r}^2_i(n) dn + \frac12 \beta \sum_{i,j=1}^2\int_0^N 	\!\!\int_0^N u({\bf r}'_i(n), {\bf r}_j(n)) dn dn' + \beta\sum_{i=1}^2 \int_0^N \phi_{ext}( {\bf r}_i(n)) dn,
	\label{eq.1}
\end{equation}
where $\ell$ is the Kuhn's length, $u({\bf r}', {\bf r})$ is the pair interaction potential , while $\phi_{ext}( {\bf r})$ is the external interaction potential.   The indexes $i,j$ stand for the two polyelectrolyte chains. Fig. \ref{fig0} schematically presents the mesoscopic model on which the present evaluation of the bridging interactions is based. Clearly the non-pairwise additive effects such as bridging between multiple macroions mediated by a single chain, have been completely disregarded in this model. The finite macroion size effects have also been disregarded. Also the model is based on a linear theory (Debye-H\" uckel) of Coulomb interactions and thus can not capture non-linearities such as charge renormalization or counterion condensation. It can however take into account the electrostatic stiffening of the chain as well as the finite chain size effects.

The grafting of the chains to their respective macroions is not taken into account explicitely on the Hamiltonian level. First of all in this model system the effects of grafting are small or indeed negligible \cite{podgornik2} to the extent that they are always overpowered by the much stronger electrostatic interactions. This would of course not be the case for {\sl e.g.} electrostatic brushes where grafting has to enter the description of the chain already at the Hamiltonian level. The grafting of the chains is only taken into account {\sl via} their center of mass coordinates in the way explained later.

\section{Variational {\sl ansatz} and formalism}

Since the interactions with external fields as well as the self-interactions along the polymer chains are highly non-linear in terms of their spatial dependence, the statistical integral corresponding to Eq. \ref{eq.1} is in general impossible to evaluate analytically and is difficult to evaluate even approximately. Instead of taking recourse to a numerical approximation we will rather introduce a harmonic {\sl variational ansatz}  \cite{kleinert1, variational} that will make the evaluation of the statistical integral straightforward. The parameters of the {\sl variational ansatz} will be chosen so as to minimize the upper bound of the exact free energy. This procedure is generally known as the {\sl Feynman - Kleinert variational method}  \cite{variational} and its application to the case of self-interacting polymer chains in the bulk  has been discussed in detail elsewhere \cite{podgornik3}. It was shown that it captures all the salient features of polyelectrolyte behavior in the bulk.

The details of the variational  approach are as follows \cite{podgornik3}: the two polyelectrolyte chains are trated as Gaussian blobs positioned at their centers of mass, ${\bf r}_{01}$ and ${\bf r}_{02}$ respectively. The width of the Gaussian blobs is determined variationally from the strength of the external fields as well as self-interactions along the chain  \cite{variational}.  The Gaussian blobs thus behave as effective "particles" with finite extensions. In this respect we can claim that the present theoretical framework represents a kind of Asakura-Oosawa theory where the effective size of the macroions is determined variationally.  The final statistical integral is then obtained by integrating over the two centers of mass ${\bf r}_{01}$ and ${\bf r}_{02}$, {\sl i.e.} the coordinates of the effective (Asakura-Oosawa) "particles",  or indeed by finding the configuration of the centers of mass that gives the largest contribution to this integral \cite{additional}.

For the variational {\sl ansatz} corrsponding to two effective Gaussian chains we will chose a general harmonic hamiltonian of the form \cite{podgornik3}
\begin{equation}
	\beta {\cal H}_0[{\bf r}_i(n)] = \frac{3}{2 \ell^2} \sum_{i=1}^2\int_0^N \dot{\bf r}_i^2(n) dn + \frac{3}{2} \sum_{i=1}^2\zeta_i^2({\bf r}_{0i}) \int_0^N \left( {\bf r}_i(n) - {\bf r}_{0i} \right)^2 dn + \beta N {\cal L}({\bf r}_{01}, {\bf r}_{02}),
	\label{eq.2}
\end{equation}
with periodic boundary conditions for ${\bf r}_i(n)$. This {\sl ansatz} is obviously still dependent on ${\bf r}_{0i}$ for $i =1,2$, that stand for the centers of mass of the two chains  {\sl i.e.} ${\bf r}_{0i} = \ul1N \int_0^N {\bf r}_i(n) dn$, as well as the functions $ \zeta_i({\bf r}_{0i})$ and ${\cal L}({\bf r}_{01}, {\bf r}_{02})$ that will be determined variationally. 

The term with $ \zeta^2_i({\bf r}_{0i})$ obviously represents an external harmonic potential, centered on ${\bf r}_{0i}$, that acts either to confine or expand the chain, depending on its sign. A simple limiting form of $ \zeta^2$ can be derived only for the case of a single self-interacting chain and is given in \cite{podgornik3}. In Eq. \ref{eq.2} this term was taken with a positive sign but we will argue later that it can as well be negative. The term $\beta N {\cal L}({\bf r}_{01}, {\bf r}_{02})$ simply represents the value of this harmonic external potential at the centers of mass of both chains. Again a  simple limiting form of $ \beta N {\cal L}$ can be derived only for the case of a single self-interacting chain \cite{podgornik3}. As will become clear when we proceed, both quantities depend in a complicated way on the interactions between the monomers as well as on the interactions between the monomers and external macroions. 

The statistical integral for the variational {\sl ansatz} can be obtained in the following form \cite{podgornik3}
\begin{equation}
	\Xi_0(N) = \int {\cal D}[{\bf r}_1(n)] \int {\cal D}[{\bf r}_2(n)] ~e^{- \beta {\cal H}_0[{\bf r}_i(n)] } = \int \!\!\int d^3{\bf r}_{01 }d^3{\bf r}_{02}~ e^{- \beta {\cal 	F}_0({\bf r}_{01}, {\bf r}_{02}}).
	\label{eq.1.1}
\end{equation}
The two polymer chains are thus represented as two effective Gaussian  "particles" with an effective Hamiltonian given by ${\cal F}_0({\bf r}_{01}, {\bf r}_{02})$. The details of the implementation of the Feynman - Kleinert ansatz  \cite{variational} for (self)interacting polymer chains have been given before \cite{podgornik3} and we will rely on the formal developments described in that work.  First of all let us introduce the radius of gyration defined as
\begin{equation}
	a_i^2({\bf r}_{0i}) = \frac{1}{3 N} \int_0^N \left< \left( {\bf r}_i(n) - {\bf r}_{0i}\right)^2 \right>  dn =  \frac{1}{3 \zeta_i} {\cal 	L}\left( \frac{\zeta_i \ell^2 N}{2} \right),
	\label{eq.1.2}
\end{equation}
where ${\cal L}(x) = \coth{x} - \ul1x $ is the standard Langevin function. One can then derive \cite{podgornik3} that minimization of the upper bound of the variational free energy with respect to the function ${\cal L}({\bf r}_{01}, {\bf r}_{02})$ leads to the following equation
\begin{equation}
	 \beta N {\cal L}({\bf r}_{01}, {\bf r}_{02}) = - \ul32  \sum_{i=1}^2 \zeta_i^2({\bf r}_{0i}) N a_i^2({\bf r}_{0i})  + \beta W({\bf r}_{01}, {\bf r}_{02}),
\end{equation}
with
\begin{equation}	
	 \beta W({\bf r}_{01}, {\bf r}_{02})) = \beta \int d^3{\bf r} ~\phi_{ext}( {\bf r}) \rho({\bf r}, {\bf r}_{01}, {\bf r}_{02}) + 
	+ \frac12 \beta  \int \!\!\int d^3{\bf r}  d^3{\bf r}'   \rho({\bf r}, {\bf r}_{01}, {\bf r}_{02}) u({\bf r}, {\bf r}')\rho({\bf r}', {\bf r}_{01}, {\bf r}_{02}).
\end{equation}
$W({\bf r}_{01}, {\bf r}_{02})$ obviously represents the total interaction free energy, due to self as well as interactions with external fields, of a smeared monomer cloud with a Gaussian density distribution.  $\rho({\bf r}, {\bf r}_{01}, {\bf r}_{02})$ stands for the combined  monomer density function of the two chains and has the form
\begin{equation}
	\rho({\bf r}, {\bf r}_{01}, {\bf r}_{02}) = \rho_{a_1^2}({\bf r}, {\bf r}_{01}) + \rho_{a_2^2}({\bf r}, {\bf r}_{02}),
\end{equation}
where for each of the chains the monomer density distribution function is given by 
\begin{equation}
	\rho_{a_i^2}({\bf r}, {\bf r}_{0i}) = \frac{N}{(2\pi a_i^2)^{3/2}} \exp{-\frac{\vert {\bf r} - {\bf r}_{0i}\vert^2}{ 2 a_i^2}} \qquad {\rm or ~else} \qquad \rho_{a_i^2}({\bf k}) = N~\exp{- \frac{k^2}{2 a_i^2}}
\end{equation}
in real and in the Fourier space, the form we will need later on. Taking now the form of the self interaction and the external fields as in Eq. \ref{eq.0.1} and  \ref{eq.0.2}  we obtain the following result
\begin{equation}
	W({\bf r}_{01}, {\bf r}_{02}) = \sum_{k=1}^2 {\cal F}_{a_1^2}\left( {\bf r}_{01} - {\bf r}_k \right) +  \sum_{k=1}^2 {\cal F}_{a_2^2}\left({\bf r}_{02} - {\bf r}_k \right) + \sum_{k=1}^2 {\cal W}_i + {\cal W}_{a_1, a_2} ({\bf r}_{01}, {\bf r}_{02}).
\end{equation}
Here ${\bf  r}_i$ stand for the position of the two macroins (to be distinguished from the position of the two centers of mass of the polymer chains ${\bf r}_{0i}$). ${\cal F}$s are due to the interaction of the chains with external fields and can be written in the form
\begin{equation}
	{\cal F}_{a_i^2}\left( {\bf r} - {\bf r}' \right) = \int \frac{d^3{\bf k}}{(2\pi)^3}  ~	\rho_{a_i^2}({\bf k}) u({\bf k}) ~e^{i {\bf k}\dot ({\bf r} - {\bf r}')}.
\end{equation}
The self-interactions of the chains correspond to the terms ${\cal W}_i$ and are given by
\begin{equation}
	{\cal W}_i = \int \frac{d^3{\bf k}}{(2\pi)^3}  ~	\vert \rho_{a_i^2}({\bf k})\vert^2 u({\bf k}),
\end{equation}
and finally the interactions between the two chains that can be derived in the form
\begin{equation}
	{\cal W}_{a_1, a_2} ({\bf r}, {\bf r}') =  \int \frac{d^3{\bf k}}{(2\pi)^3}  ~	\rho_{a_1^2}({\bf k}) u({\bf k}) \rho_{a_2^2}(-{\bf k})~e^{i {\bf k}\dot ({\bf r} - {\bf r}')}.
\end{equation}
This is all straightforward generalization of the variational theory set up previously for a single chain \cite{podgornik3}. The functions $\zeta_i({\bf r}_{0i})$ are next obtained by minimizing the upper bound to the exact free energy with respect to $a_i^2$ leading to
\begin{equation}
	\ul32  \zeta_i^2({\bf r}_{0i}) N  = \beta \frac{\partial}{\partial a_i^2} W({\bf r}_{01}, {\bf r}_{02}).
	\label{eq.3.1}
\end{equation}
The effective centers-of-mass free energy of the two polymer chains is finally given by the expression
\begin{eqnarray}
	\beta {\cal F}_0({\bf r}_{01}, {\bf r}_{02}) &=& 3 \sum_{i=1}^2 \log{\frac{\sinh{\frac{\zeta_i \ell N}{2}}}{\frac{\zeta_i \ell N}{2 }}} - \ul32 \sum_{i=1}^2  \zeta_i^2 N a_i^2  + \beta W({\bf r}_{01}, {\bf r}_{02}) = \nonumber\\
	&=& \beta {\cal F}_2({\bf r}_{01}, {\bf r}_{02}) + \beta W({\bf r}_{01}, {\bf r}_{02}),
	\label{eq.3}
\end{eqnarray}
where we separated out the harmonic part of the free energy $ \beta {\cal F}_2({\bf r}_{01}, {\bf r}_{02})$. The first  two terms of this variational free energy represent the entropy of the Gaussian chain and the last  one is due to the interactions with the external fields and self-interactions. 
These are the basic equations of the Feynman - Kleinert variational theory as applied to the self interacting polyelectrolyte chains. They are still quite complicated because of the dependence on the center-of-mass coordinates ${\bf r}_{0i}$ and the final integration over these variables in Eq. \ref{eq.1.1}. 

If there are no external fields that break the translational symmetry of the problem  it can be easily seen \cite{podgornik3} that the dependence on ${\bf r}_{0i}$ vanishes and the solution of the variational equations is straightforward. With external fields the final quite complicated ${\bf r}_{0i}$ integration can be obtained only numerically. In the case that ${\cal F}_0({\bf r}_{01}, {\bf r}_{02})$ scales with a positive power of $N$ and $N$ is large enough, there is however an additional quite accurate approximation to circumvent this final integration \cite{additional}. It consists of the saddle point evaluation of the final integration with respect to  ${\bf r}_{0i}$, that is of an additional minimization of  ${\cal F}_0({\bf r}_{01}, {\bf r}_{02})$ with respect to ${\bf r}_{01}$ as well as ${\bf r}_{02}$. This means that Eq. \ref{eq.1.1} can be written in an approximate form
\begin{equation}
	\Xi_0(N) =  \int \!\!\int d^3{\bf r}_{01 }d^3{\bf r}_{02}~ e^{- \beta {\cal 	F}_0({\bf r}_{01}, {\bf r}_{02})} \approx e^{- \beta {\cal F}_0({\bf r}^*_{01}, {\bf r}^*_{02})}.
	\label{eq.4.1}
\end{equation}
where ${\bf r}^*_{01}, {\bf r}^*_{02}$ are given as solutions of the saddle-point condition
\begin{equation}
	\frac{\partial {\cal F}_0({\bf r}^*_{01}, {\bf r}^*_{02})}{\partial {\bf r}^*_{01}} = \frac{\partial {\cal F}_0({\bf r}^*_{01}, {\bf r}^*_{02})}{\partial {\bf r}^*_{02}} = 0.	\label{eq.4}
\end{equation}
Thus we obtain a simple explicit and accurate estimate for the free energy of two self-interacting polyelectrolyte chains in an external field, {\sl viz.} ${\cal F} = - kT~\Xi_0(N) \approx  {\cal F}_0({\bf r}^*_{01}, {\bf r}^*_{02})$. In what follows we will always assume that all the solutions of the variational equations have to be symmetric with respect to the two chains. There is no reason on the pair-potential level to assume otherwise.

Since the solution of variational equations is in general quite complicated we give here a little preview of what exactly we will be calculating in what follows. We will show that {\sl grosso modo} the solution of the variational problem has two branches depending on the relative magnitudes of the interaction with external fields and the self and mutual interaction of the chains. The two branches of the solution are:
\begin{itemize}
\item what we call a {\sl strong coupling branch} that corresponds to  $\zeta_i^2({\bf r}_{0i}) > 0$ in the variational equation Eq. \ref{eq.1.1}, and thus to the dominance of the interactions of the chains with the external macroion fields, the self- and mutual interactions of the chains being a small perturbation. The strong coupling branch furthermore bifurcates into two different sub-branches depending  on the solution of the additional minimization implied by Eq. \ref{eq.4}: in the {\sl weakly paired} sub-branch the chains are associated each with its own grafting macroion and in the {\sl strongly paired} sub-branch both chains share the two macroions on the average.
\item and what we term a {\sl weak coupling branch} where  $\zeta_i^2({\bf r}_{0i}) < 0$ and thus corresponds to the case where the self and mutual interactions of the chain are dominant and the interactions with external macroion fields are perturbative. {\sl Coupling} in both cases thus refers to coupling with the external macroion field.
\end{itemize}
In both cases we can in general observe some bridging effects but they are several orders of magnitude stronger in the first case. Nevertheless they are always present to some extent. After this introductory survey of the nature of the solutions of the variational problem, we are ready to find these solutions explicitly. 

A note on the grafting of the chain is in order at this point. Both in the {\sl weakly paired} as well as the {\sl strongly paired} states the electrostatic adsorption energy more than the grafting itself determines the statistics of the chain. Grafting the chains, by fixing {\sl e.g.} ${\bf r}_i(0)$ to be at the surface of the macroion, would change none of the conclusions reached below, provided of course that the size of the macroions is small compared to the size of the chains and that we have only one chain associated with each of the macroions. It would however introduce some serious complications into our formalism thus obscuring its straightforward interpretation.

\section{Solution of the variational equations}

We are now ready to solve the general variational equations for the case of two polyelectrolyte chains with two external point macroions.   As we already announced we consider only symmetric solutions for which $a_1^2 = a_2^2 = a^2$, but in general with ${\bf r}_{01} \neq {\bf r}_{02}$. This symmetrization will be applied to the results derived below in their final form.

A little straihgtforward algebra then leads to the following form of the total variational free energy Eq. \ref{eq.3} 
\begin{eqnarray}
	\beta {\cal F}_{a_i^2}\left( {\bf r} - {\bf r}' \right) &=& - \frac{\ell_B M N 2^{3/2}}{\pi  a} 	 f_1\left(  \frac{\sqrt{2}}{a} \vert {\bf r} - {\bf r}' \vert, \frac{\kappa a}{\sqrt{2}}\right) \nonumber\\
	\beta {\cal W}_i &=& \frac{\ell_B N^2}{\pi a} f_1(0, \kappa a) \nonumber\\
	\beta {\cal W}_{a_1, a_2} ({\bf r}, {\bf r}') &=&  \frac{4 \ell_B N^2}{\pi a} f_1\left( \frac{1}{a}  \vert {\bf r} - {\bf r}' \vert, \kappa a \right).
\end{eqnarray}
$\ell_B = \frac{e_0^2}{4\pi \epsilon\epsilon_0 kT}$ was introduced above as the Bjerrum length. Also we defined the following function
\begin{eqnarray}
	f_{\lambda}(y,t) &=& \int_0^\infty \frac{u \sin{u y} ~e^{-\lambda u^2}}{y (u^2 + t^2)} du = 	\nonumber\\
	&=& \frac{\pi}{4 y}~e^{\lambda y^2} \left( 2 e^{-y t} - e^{-y t} Erfc(\frac{y}{2 \sqrt{\lambda}} - t \sqrt{\lambda}) - e^{y t} Erfc(\frac{y}{2 \sqrt{\lambda}} + t \sqrt{\lambda})\right),
\end{eqnarray}
where $Erfc(x)$ is the standard complementary error function. On the other hand the variational equation Eq. \ref{eq.3.1} can be obtained just as straihgtforwardly  as,
\begin{equation}
	\beta \frac{\partial}{\partial a_1^2} W({\bf r}_{01}, {\bf r}_{02})  = \frac{\ell_B N}{\pi a^3} \left[ 	2^{3/2} M  \sum_{k=1}^2	g \left(\frac{\sqrt{ 2}}{a} \vert {\bf r}_{01} - {\bf r}_k\vert, \frac{\kappa a }{\sqrt{2}}\right) - 2 N g \left(\frac{\vert {\bf r}_{01} - {\bf r}_{02}  \vert}{a} , \kappa a \right)
	- N g \left(0, \kappa a \right) \right].
	\label{eq.4.0.1}
\end{equation}
A similar equation could be obtained also for $\beta \frac{\partial}{\partial a_2^2} W({\bf r}_{01}, {\bf r}_{02})$ except that ${\bf r}_{01}$ on the r.h.s. would be turned into ${\bf r}_{02}$. The following new function  was defined above
\begin{equation}
	g(y,t) = - \frac{\partial}{\partial \lambda}f_{\lambda}(y, t) \vert_{\lambda=1}.
\end{equation}
What the variational equation Eq. \ref{eq.4.0.1} really asserts is which terms are important in determining the statistical conformation of the chain, {\sl i.e.} $a^2_i$ in our case. The first term on the r.h.s. of Eq. \ref{eq.4.0.1} is due to the interactions with the macroions, the second one is due to the interactions between the two chains and the last one is the self interaction of the chains. The conformation of the chain as descibed by $a^2_i$ is thus determined by the relative magnitudes of these three terms.

The properties of the solution of the variational equation Eq. \ref{eq.4.0.1} first of all depend crucially on the sign of the r.h.s. of the Eq. \ref{eq.4.0.1}, thus on the fact whether $ \zeta^2$ is positive or negative. The sign of this term tells for each value of the separation between the macroions  whether it is the interactions with the external fields or the self-interactions of the chain that determine the statistical conformation  of the chain. 

In view of the form of the variational {\sl ansatz} Eq. \ref{eq.2} the positive sign corresponds to a general confinement of the chain if compared to the case with no interactions. We will refer to the ensuing interactions between the two macroions mediated by the polyelectrolyte chain as {\sl strong coupling} limit. In the opposite case the chain is expanded if compared to the case with no interactions, and we will refer to the ensuing polyelectrolyte mediated interactions as {\sl weak coupling} limit. Both terms will be explained further below. The final closure for this system of variational equations is provided by the relation between $\zeta$ and $a$, Eq. \ref{eq.1.2}. 

Since in this model the external macroions break the translational symmetry of the system we apply also the minimization condition Eq. \ref{eq.4} with respect to ${\bf r}_{01}$ as well as  ${\bf r}_{02}$, in order to avoid the final complicated integral over the centers-of-mass of the two polyelectrolyte chains. This minimization introduces additional features of the solutions of the variational equations. Taking into account the Gaussian-like form of the function $f_{1}(y,t)$ we realize, that there are in fact two different symmetric solutions to equation Eq. \ref{eq.4}: 
\begin{itemize}
\item ${\bf r}_{01} = {\bf r}_1$ and  ${\bf r}_{02} = {\bf r}_2$, {\sl i.e.} each of the chain remains associated with its grafting macroion
\item ${\bf r}_{01} = {\bf r}_{02} = \ul12 ({\bf r}_1 + {\bf r}_2)$, {\sl i.e.} each chain is shared by the two macroions symmetrically.
\end{itemize}
Here we assumed that the first chain is grafted to the first macroion while the second one is grafted to the second macroion. We refere to the configuration of the polyelectrolye chains in the first case as {\sl weakly paired} and in the second case as {\sl strongly paired}. The terms are self-explanatory: in the first two cases the chain is confined to one of the macroions, whereas in the second case it is confined or bound by both of them. A schematic representation of the solutions of the variational equations is presented on Fig. \ref{fig0}.

\section{Strong coupling limit}

Once again {\sl strong coupling} means that external fields dominate the statistical configuration of the chain and thus $\zeta^2 > 0$. In this domain of the parameter space the effect of the interactions of the polyelectrolyte chain with  the macroions, the term proportional to $M$ in Eq. \ref{eq.4.0.1},  determines the overall configuration of the chain. 

The strong coupling limit entails however two different polyelectrolyte equilibrium states as discussed above, depending on the minimization with respect to ${\bf r}_{01}, {\bf r}_{02}$: the first state, stable for small values of the separation between the macroions, is due to the {\sl strong pairing} configuration of the chain with ${\bf r}_{01} = {\bf r}_{02} = \ul12 ({\bf r}_1 + {\bf r}_2)$. The variational equation for $\zeta$ in this case reads
\begin{equation}
	\ul32 \zeta^2 = \frac{\ell_B N}{\pi a^3} \left[ 2^{5/2} M g \left(\frac{\sqrt{ 2}}{2 a} \vert {\bf r}_1 - 	{\bf r}_2\vert, \frac{\kappa a }{\sqrt{2}} \right) 
	- 3 N g \left(0, \kappa a \right) \right],
	\label{eq.4.0.0}
\end{equation}
while the corresponding free energy has the form
\begin{eqnarray}
	\beta {\cal F}_0 &=& 2 \beta {\cal F}_2(\vert {\bf r}_1 - {\bf 	r}_2\vert) - \nonumber\\
	& & - \frac{\ell_B N}{\pi a} \left(  2^{7/2} M 	\left[  	f_1 \left(\frac{\sqrt{ 2}}{2 a} \vert {\bf r}_1 - 	{\bf r}_2\vert, \frac{\kappa a }{\sqrt{2}}\right)  \right] - 6 N f_1 \left(0, \kappa a \right) \right).
	\label{eq.4.1.1}
\end{eqnarray}
The form of the dependence ${\cal F}_2(\vert {\bf r}_1 - {\bf 	r}_2\vert) $ is of course given implicitely {\sl via} the dependence of $\zeta$ and $a$. Once again the chain here is bound to both macroions and its statistical properties are dominated by the interaction with the charges on the macroions. One would expect that the polyelectrolyte mediated interactions between the macroions would be strongest in this case. Obviously for large enough $\vert {\bf r}_1 - {\bf r}_2\vert$  the r.h.s. of Eq. \ref{eq.4.0.0} can become negative, going first through zero. This is due to the fact that $g(y,t)$ is a decaying function of $y$. At this point the above solution ceases to be  stable and we have a transition from the {\sl strongly paired} to {\sl weakly paired} branch of the strong coupling limit. The transition depends on the macroion parameters such as the magnitude of their charges as well as the length of the chains. In this sense it represents a finite size (of the chains) effect.

The {\sl weakly paired} configuration is characterized by ${\bf r}_{01} = {\bf r}_1$ and  ${\bf r}_{02} = {\bf r}_2$ and is the stable branch at larger separations between the macroions.  Here the variational equation for $\zeta$ becomes 
\begin{equation}
	\ul32 \zeta^2 = \frac{\ell_B N}{\pi a^3} \left[ 2^{3/2} M 	\left[  	g \left(0, \frac{\kappa a }{\sqrt{2}}\right) + 	g \left(\frac{\sqrt{ 2}}{a} \vert {\bf r}_1 - {\bf r}_2\vert, 	\frac{\kappa a }{\sqrt{2}}\right) \right] - 2 N g \left(\frac{\vert {\bf r}_1 - {\bf r}_2\vert}{a} , \kappa a 	\right) - N g \left(0, \kappa a \right) \right],
	\label{eq.4.2}
\end{equation}
The corresponding free energy in this case can be obtained as
\begin{eqnarray}
	\beta {\cal F}_0 &=& \beta {\cal F}_2(\vert {\bf r}_1 - {\bf 	r}_2\vert) - \nonumber\\
	& & - \frac{\ell_B N}{\pi a} \left(  2^{5/2} M 	\left[  f_1 \left(0, \frac{\kappa a }{\sqrt{2}}\right) +	f_1 \left(\frac{\sqrt{ 2}}{a} \vert {\bf r}_1 - {\bf r}_2\vert, \frac{\kappa a }{\sqrt{2}}\right)  \right] - \right.\nonumber\\
	& & \left. - 4 N f_1  \left( \frac{\vert {\bf r}_1 - {\bf r}_2\vert }{a} ,\kappa a \right)     	- 2 N f_1 \left(0, \kappa a \right) \right).
	\label{eq.4.2.1}
\end{eqnarray}
The most important term to determine the conformation of the chain is the interaction with the single macroion and is thus only weakly dependent on the separation between them. These are the first and the last term in the r.h.s. of Eq. \ref{eq.4.2}. The separation dependent terms act only as a perturbation to these terms. It is thus to be expected that the polyelectrolyte mediated interactions will be much weaker in this case. 

Also in order to get the {\sl interaction} free energy in the {\sl weakly paired} configuration one needs in addition to subtract  the terms that do not depend on the separation between the macroions from the total free energy. This is the standard way to get the interaction free energy. Again the form of the dependence ${\cal F}_2(\vert {\bf r}_1 - {\bf r}_2\vert) $ is given implicitely {\sl via} the dependence of $\zeta$ and $a$.

\begin{figure}[ht]
\begin{center}
    \epsfig{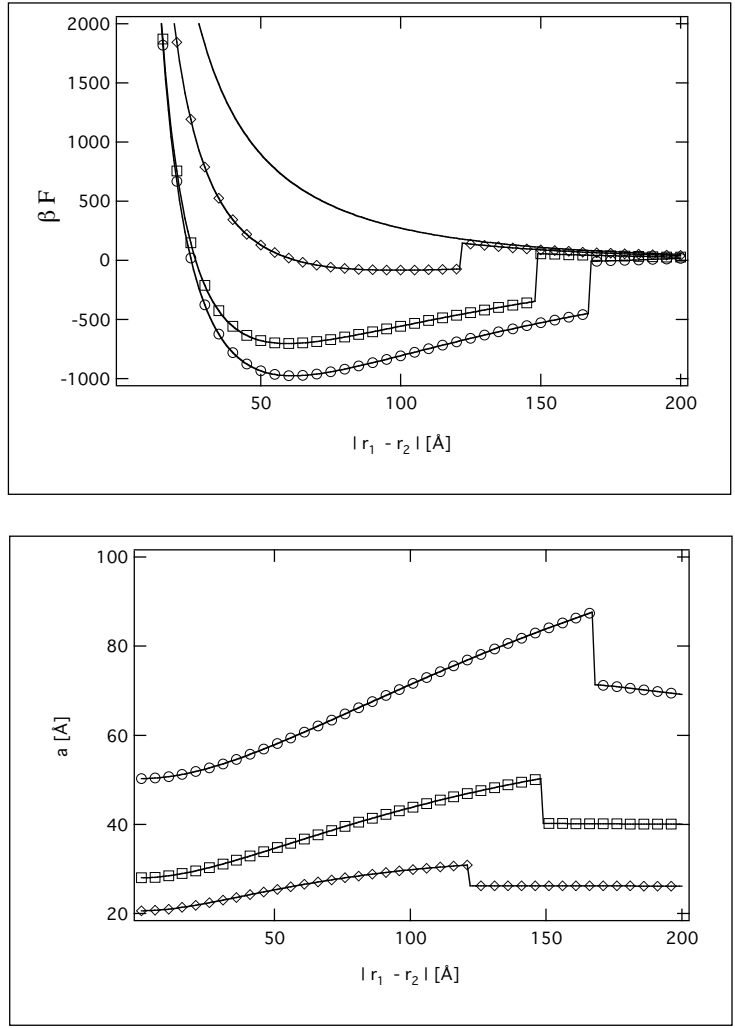}
\end{center}    
\caption{ $\beta {\cal F}_0$ and $a$ in the strong coupling limit. The upper graph shows the dependence of $\beta {\cal F}_0$ for $M = 100$ for $N = 10 $ (losenge), $30$  (square), $100$  (circle) at $1 mM$  on the separation between the macroions $\vert {\bf r}_1 - {\bf r}_2\vert$. The lower graph shows the dependence of the size of the chain $a$ on the separation between the macroions for the same values of parameters.The length of the chain $N$ obviously determines the separation between macroions where the snapping of the chain between the strongly paired and weakly paired states occurs. In the strongly paired configuration we have a well developed regime of attractive bridging interactions leading to a an effective screened coulomb repulsion in the weakly paired regime. In all cases $\ell = 10 $. The bold line represents the pure Debye-H\" uckel interactions between the macroions. }
\label{fig1}
\end{figure}

The form of the solution of Eq. \ref{eq.4.1} and Eq. \ref{eq.4.2} as well as the corresponding polyelectrolyte mediated interaction free energy is presented in Fig.\ref{fig1}. We see that at small enough separations the chain is in the {\sl strongly paired} configuration, being confined symmetrycally  by both macroions. In this regime the external field trying to confine the chain to both macroions wins over the chain entropy that is expanding the chain. The entropy of the parts of the chain spanning the region between the macroions is quite low. Its size, as described by $a$, in this case depends monotonically on the separation between the macroions that are effectively stretching it. 

\begin{figure}[ht]
\begin{center}
    \epsfig{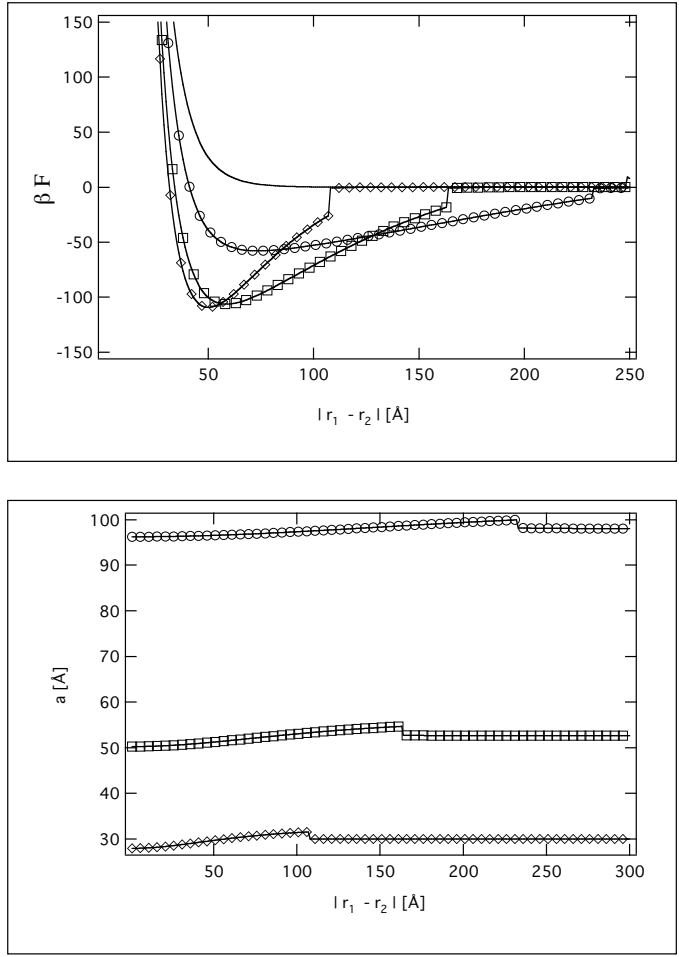}
\end{center}    
\caption{ $\beta {\cal F}_0$ and $a$ in the strong coupling limit. The upper graph shows the dependence of $\beta {\cal F}_0$ for $M = 100$ for $N = 10, 30, 100$ at $60 mM$ on the separation between the macroions $\vert {\bf r}_1 - {\bf r}_2\vert$.  The lower graph shows the dependence of the size of the chain $a$ on the separation between the macroions for the same values of parameters. The bold line represents the pure Debye-H\" uckel interactions between the macroions.}
\label{fig2}
\end{figure}

At the instability point, reached at a well defined value of the separation between the macroions, the chain entropy scores a partial victory over the interactions with the macroions forcing the chain to remain close to the macroion to which it is grafted. At this transition the chain basically relaxes the low entropy configurations of its parts confined between both macroions by snapping back to macroion to which it is grafted. After that the size of the chain is basically determined solely through the interactions of the chain with its grafting macroion and remains constant with separation between the macroions. These conclusions reached on the basis of the two chain variational approach are very similar to the  existing simulation data \cite{podgornik2} for one chain in the field of two macroions.

This scenario of chain conformations is clearly illustrated in Figs. \ref{fig1} and \ref{fig2} where one can follow the transtition of the chain from the {\sl strongly paired} to the {\sl weakly paired} configuration  {\sl via} the changes of $a$ as a function of the separation between the macroions for two different values of the ionic strength of the uni-valent salt solution, {\sl viz} $1$ mM and $60$ mM. Clearly the overall effect of the salt is to quench the magnitude of the bridging interaction. We note about one order of magnitude difference in the strength of the bridging attraction at both salt activities. The increase in salt activity also quenches the difference between the {\sl weakly paired} and the {\sl strongly paired} configuration. If we compare the radius oif gyration of the chain (lower graphs on Figs. \ref{fig1} and \ref{fig2}) we see that at higher salt the separation between the macroions has a smaller effect on the size of the chain.

The ensuing chain mediated interaction free energy follows closely the equilibrium configuration of the chain between the macroions. For a {\sl strongly paired} chain the interaction free energy shows a pronounced attractive contribution stemming from the coupling $M \times N$ term in  Eq. \ref{eq.4.1.1}. At the point where the chain snaps from the {\sl strongly paired} to the {\sl weakly paired} configuration there is also a corresponding jump in the free energy due to  much less pronounced chain mediated interactions. It is interesting to analyze the asymptotic behavior of the interaction free energy Eq. \ref{eq.4.2.1} in the {\sl weakly paired} branch of the solution. Expanding $f_1(y,t)$ for large values of $y$ we see that all the chain dependent parts of the free energy finally just renormalize the direct Debye-H\" uckel interactions between the macroions . Thus asymptotically instead of a Debye-H\" uckel interaction of strength proportional to $M^2$ we simply end up with its strength being proportional to $(M-N)^2$. The chain snapped back to the grafting macroion and simply renormalized its charge.

There is one interesting remark that we can make here. Gurovitch and Sans \cite{gurovitch} studied polyelectrolyte adsorption of a single chain to a (point) charged macoion. Their case thus correspond to an {\sl weakly paired} branch at infinite separation between the macroions in our terminology, that would correspond to Eq, \ref{eq.4.2} with $\vert {\bf r}_1 - {\bf r}_2\vert \longrightarrow \infty$ leading to
\begin{equation}
	\ul32 \zeta^2 = \frac{\ell_B N}{\pi a^3} \left[ 2^{3/2} M g \left(0, \frac{\kappa a }{\sqrt{2}}\right)  - N g \left(0, \kappa a \right) \right].
\end{equation}
Clearly in the vanishing salt limit $\kappa a \longrightarrow 0$, which is in fact the case treated in \cite{gurovitch}, the polyelectrolyte chain can adsorb until its charge (or the number of monomers) becomes equal to $N = 2^{3/2} M$, there is thus maximal overcharging in the amount of   $2^{3/2} \approx 2.83$ which is indeed very close to the value derived by them by a completely different method {\sl viz.} $ 15/6 = 2.5$. Though the approach of these authors has been criticized \cite{gurovitch} our results are more then consistent with theirs. More could be said on polyelectrolyte adsorption and overcharging \cite{stoll} but we will focus here strictly on the interaction {\sl i.e.} bridging aspects of the problem.

\section{Weak coupling limit}

In this case the effect of electrostatic self-interaction of the chain, the term proportional to $N$ in Eq. \ref{eq.4},  determines the overall configuration of the chain.  If the effect of the external fields would be indeed negligible we have shown in a previous publication \cite{podgornik3} that the electrostatic interactions would stiffen up the chain and give it a rod-like appearance quantified by the scaling $ a \sim N$. We expect that even with external fields originating at the macroions the chain will essentially assume this type of extended configurations in this limit, modified by the perturbative effect of both macroions. Simulations of single chain adsorption \cite{stoll} are completely consistent with this picture since for large $N$ protruding rodlike tails are observed that correspond to electrostatically stiffened portions of the chain. Calculations of Nguyen and Shklovskii \cite{nguyen}  also lead to the same qualitative picture of chain adsorption in this limit. Again here we are not interested in adsorption {\sl per se} but rather in the bridging effects in interaction between the macroions, so we will skip the detailed comparison of our work with polyelectrolyte adsorption studies.
 
With the external fields present the stiffening of the chain depends on the details {\sl i.e.} strength of the term proportional to $M$ in the variational equation Eq. \ref{eq.4} and is difficult to quantify in scaling terms, unless the effect of the macroions is thoroughly negligible, which is not the situation we are trying to investigate.
\begin{figure}[ht]
\begin{center}
    \epsfig{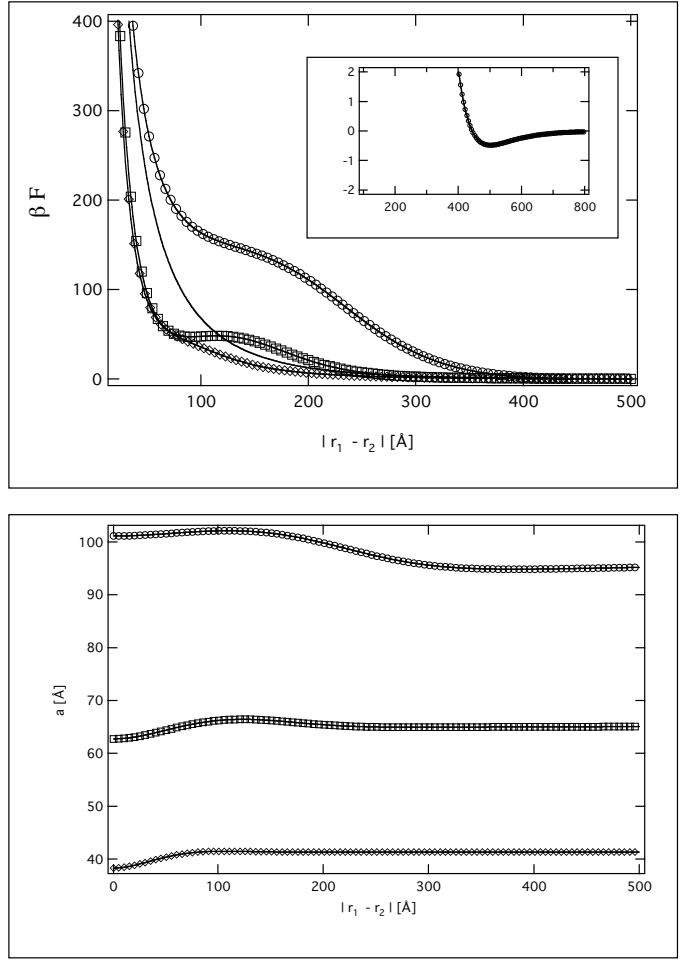}
\end{center}    
\caption{$\beta {\cal F}_0$ and $a$ in the weak coupling limit for $M = 40$ and $N = 50, 20, 10$ at $1 mM$.  There exists only an {\sl weakly paired} state in this case and the effect of the external fields of the macroions is much less pronounced then in the strong coupling limit. The length of the chain $N$ determines primarily the steric repulsion effect due to the interpenetration of stiffened grafted chains, the "coronas" of both macroions, on close approach.  The inset shows the residual very weak coupling interaction at large separations between macroions with $N = 100$. The size of the chain tends to grow slightly on approach of the macroions because of the interpenetration of the "coronas" of both macroions. The bold line represents the pure Debye-H\" uckel interactions between the macroions.}
\label{fig3}
\end{figure}
Since here the effect of the macroions is small the polyelectrolyte chains can never be {\sl strongly paired} by both macroions. We thus remain solely with  the {\sl weakly paired} configuration of the chains due to the grafting to the macroions. The solution of the variational equation Eq. \ref{eq.4} thus only has one branch in this case and is given by
\begin{equation}
	\ul32 \zeta^2 = \frac{\ell_B N}{\pi a^3} \left[   2 N g \left(\frac{\vert {\bf r}_1 - {\bf r}_2\vert}{a} , \kappa a 	\right) + N g \left(0, \kappa a \right) - 2^{3/2} M 	\left[  	g \left(0, \frac{\kappa a }{\sqrt{2}}\right) + 	g \left(\frac{\sqrt{ 2}}{a} \vert {\bf r}_1 - {\bf r}_2\vert, 	\frac{\kappa a }{\sqrt{2}}\right) \right]\right],
	\label{eq.5.1}
\end{equation}
Clearly this equation is obtained by the substitution $\zeta \longrightarrow i \zeta$ from Eq. \ref{eq.4}. This transformation should be taken into account also in Eq. \ref{eq.1.2} leading to
\begin{equation}
	a^2 =   \frac{1}{3 \zeta} {\cal L}'\left( \frac{\zeta \ell^2 N}{2 \sqrt{3}} \right),
	\label{eq.1.2'}
\end{equation}
where now ${\cal L}'(x) = \ul1x - \cot{x}$.  The interaction with the macroion, the $M$ term in the Eq. \ref{eq.5.1}, can modify the value of the size of the chain but it has no effect any more on the stability of the solution. The numerical solutions to Eq. \ref{eq.5.1} are presented in Figs. \ref{fig3} and \ref{fig4}. Clearly the size of the polyelectrolyte chain in this case shows no discontinuities though it is still, to a lesser extent than before, effected by the positions of the two macroions. The term {\sl weak coupling} thus seems appropriate for the behavior of the chain in this region of the parameter space.
\begin{figure}[ht]
\begin{center}
    \epsfig{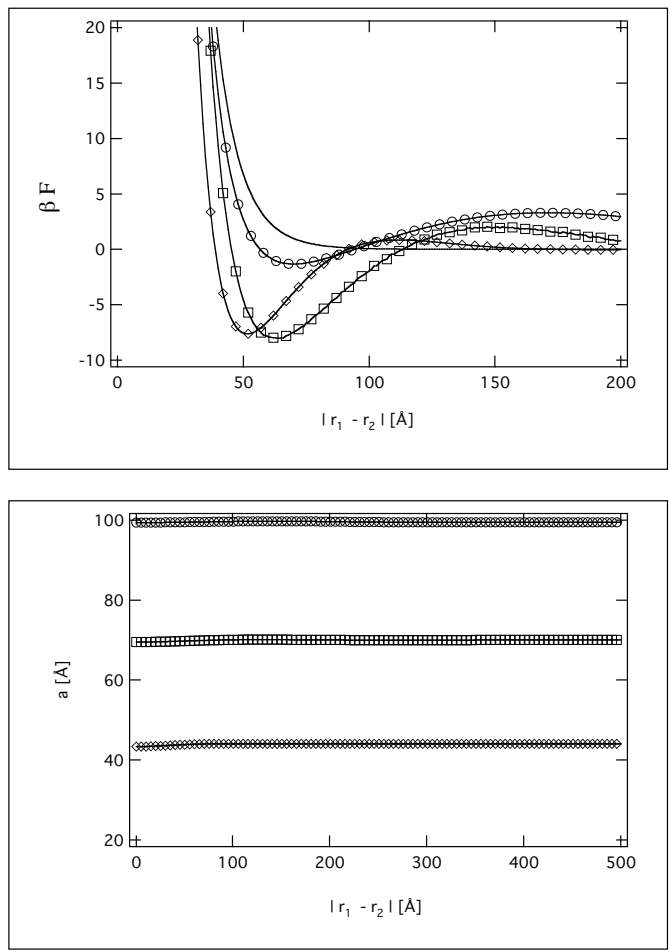}
\end{center}    
\caption{$\beta {\cal F}_0$ and $a$ in the weak coupling limit  for $M = 40$ and $N = 50, 20, 10$ at $60 mM$. Clearly again the overall effect of the salt is to quench the effects of the macroion fields. There is weak residual bridging at very small separation now, note the scale of the interactions energy, that goes smoothly into a weak "corona" repulsion at larger separations. The details of both the residual bridging as well as "corona" repulsions depend on the values of $M$ and $N$. The bold line represents the pure Debye-H\" uckel interactions between the macroions.}
\label{fig4}
\end{figure}

The free energy  is now given by an equation similar to Eq. \ref{eq.4.2.1} but with the change $\zeta \longrightarrow i \zeta$ well taken into account in ${\cal F}_2(\vert {\bf r}_1 - {\bf r}_2\vert)$. It leads to the following result
\begin{eqnarray}
	\beta {\cal F}_0 &=& 6 \log{\frac{\sin{\frac{\zeta \ell N}{2 }}}{\frac{\zeta \ell N}{2 }}} + 3~ \zeta^2 N a^2 - \nonumber\\
	& & - \frac{\ell_B N}{\pi a} \left(  2^{5/2} M 	\left[  f_1 \left(0, \frac{\kappa a }{\sqrt{2}}\right) +	f_1 \left(\frac{\sqrt{ 2}}{a} \vert {\bf r}_1 - {\bf r}_2\vert, \frac{\kappa a }{\sqrt{2}}\right)  \right] - \right.\nonumber\\
	& & \left. - 4 N f_1  \left( \frac{\vert {\bf r}_1 - {\bf r}_2\vert }{a} ,\kappa a \right)     	- 2 N f_1 \left(0, \kappa a \right) \right).
	\label{eq.4.3.1}
\end{eqnarray}
Again because the solution of the variational equations here remains on a single branch all the time,  showing no jump from one stable branch to another one ("snapping" of the chain). The free energy shows no discontinuities either, though it still depends on the separation between   macroions. 

This state of affairs introduces new features in the polyelectrolyte mediated interactions. First of all there is a clearly discernible, see Fig. \ref{fig3}, {\sl repulsion} at smaller separations. It is due to the interpenetrating "coronas", {\sl i.e.} extended configurations of the grafted chains, on approach of the macroions. If there would be many chains grafted to both macroions this incipient repulsion would develop into a full-blown brush repulsion regime. Since we have only one chain per macroion the effect of "coronal" interpenetration is rather weak, but nevertheless clearly discernible. Its range depends on the size of the chain $N$ as well as the amount of salt which regulates the overall extension of the chain. Fig. \ref{fig4} clearly shows that salt quenches the "coronal" repulsion.

It is only at larger separations, see the inset of Fig. \ref{fig3}, that residual, indeed very weak coupling, is finally discerned. The electrostatically extended chains can still make weak bridges to the other macroions but since this can happen only at sufficiently large separations, the ensuing bridging is much attenuated. On adding the salt this effect is displaced towards smaller separations because the extent of the chain is deminished by the salt as well. For larger ionic strengths we are thus left with weak coupling at smaller separations, see Fig. \ref{fig4}, where it is nevertheless stronger than in small salt, compare again inset to Fig. \ref{fig3}. Bridging, no matter what its source, is of course stongly dependent on the separations between the macroions and is in general stronger for smaller separations.  Clearly the weak coupling interaction in the case of large salt resembles much more  "sticky macroions" then (relatively) long range bridging interaction.

In the weak coupling limit there is thus an additional feature stemming from the polyelectrolyte mediated interactions which is due to "coronal" interpenetration and marks the incipient brush repulsion that would be developed fully if more chains were grafted to each macroions. The (weak) bridging attraction in this case is overall small and is in constant competition with "coronal" interpenetration interactions. 

\section{Discussion}

The polyelectrolyte bridging interaction analyzed here is obviously very rich in its features and depends crucially on the region of the parameter space under investigation. We showed that attractive bridging interactions effectively come in two varieties: the strong coupling, where interactions between the chains and the macroions are dominant, and the weak coupling, where self-interactions of the chains are dominant. Bridging interaction can be obtained in both cases but is a couple of orders of magnitude larger in the {\sl strong coupling} limit. It the {\sl weak coupling} limit the attractive bridging interactionis are overwhelmed by the incipient electrostatic repulsions between the overlapping polyelectrolyte "coronas" as well. In both cases however the effective polyelectrolyte mediated interaction potential is strongly non-monotonic and shows pronounced variation with respect to the length of the polyelectrolyte chains and the screening length of the underlying electrostatic interactions.

The variational formulation of the bridging interaction problem which lies at the basis of our approach, has several advantages as well as drawbacks. The main feature of the formalism is that it allows for the transition between the strong pairing and weak pairing states of the polyelectrolyte chains which is clearly a finite size effect and would thus be missed on the ground-state dominance level. The latter has been used succesfully for the polyelectrolyte mediated interactions between macroscopic surfaces with intervening long polyelectrolyte chains \cite{podgornik1,borukhov}. Another feature of the variational formulation is that it can describe the snapping of the chain at large enough macroion separations which is the most important feature of a finite chain size effect. We need to reiterate again that this phenomenon is absent in the ground state dominance {\sl ansatz}. The snapping of the chain from the configuration where it is partitioned between both macroions to one, where it is adsorbed to a single one, is of course due to an interplay and balance between chain adsorption energy,  chain self-interaction energy and configurational entropy of the chain. The balance depends on the size of the chain and the separation between the macroions and leads to an abrupt transition between the two configurations that has a well discerned imprint also on the polyelectrolyte mediated interaction between the macroions. In general one sees bridging only for chain cofigurations where it is partitioned by the two macroions symmetricaly, {\sl i.e.} in what we dubbed the strong pairing configuration. 

The attractive bridging interaction is typicaly about a hundred times stronger in the {\sl strong coupling} limit if compared to its {\sl weak coupling} counterpart (whence the designation of the two limiting cases). This is intuitively quite easy to grasp, since one can expect strong coupling to emerge only when the interactions between the chain and the macroions dominate the statistical properties of the system. If however the dominant interactions in the system are self-interactions of the chains, the polyelectrolytes clearly mediate only insignificantly the interactions between the macroions. In this case a much more important feature of the interaction is the interpenetration of the polyelectrolyte "coronas", that can in some cases lead to pronounced repulsions between "dressed" macroions. These repusions would clearly stabilize the macroion interactions.

The main drawback of the present analysis of the bridging interaction problem, apart from it being a purely pairwise additive formulation, is the linearized (Debye-H\" uckel) form of electrostatics. All non-linear effects are thus {\sl a priori} excluded. On this level the main effect of the salt is to attenuate the bridging interaction as well as the repulsive interaction between the polyelectrolyte "coronas". In this respect the variational approach is inferior to the ground-state dominance ansatz. One possible way out would be to formulate also the electrostatic part of the problem on a variational level where all the Debye-H\" uckel parameters would be determined self-consistently. We leave this exercise for future work.

Another importan omission of our method is the size of the macroions that does not feature explicitly in our formulation. The finite size of the chain in weakly paired or strongly paired configuration clearly showed by our numerical results (see the lower graphs in Figs. \ref{fig1} and \ref{fig2}) is thus not due to the finite size of the adsorbing macroions as in more realistic simulations \cite{podgornik2, stoll} but is an entropy-energy competiton effect: high adsorption energy vs. low configurational entropy in the weakly paired state, leading to the finite size of the weakly paired state even with a point adsorbing macroion. The omission of the finite size of the macroion tends to overestimate the polyelectrolyte mediated interactions and underestimate the size of the chain in the weakly paired as well as strongly paired configurations. This type of finite size effects can be straightforwardly incorporated into the statistics of free, non-interacting chains \cite{borisov}, but would be unfortunately difficult to incorporate into the Feynman-Kleinert variational method and were thus ignored in our formulation.  Alternative approaches would thus have to be considered \cite{netz}.

At this stage it does not seem reasonable to compare the second virial coefficient derived from our calculation of the effective pair interactions with the experiment on NCPs \cite{raspaud}. There would be just too many adjustments that one would have to put in by hand but that would have a crucial effect on the ensuing numerical results: the effective charge of the macroion due to non-linear Poisson-Boltzmann effect, the effective charge of the chains which depends strongly on the local ionic equilibrium of the dissociable amino acid groups, the effective length of the chains that are free enough to behave as flexible polyelectrolytes etc. Nevertheless, if one chooses to ignore all these additional complications the virial coefficient in the {\sl strong coupling} limit comes out always monotonically (decreasing) dependent on the screening length, {\sl i.e.} ionic strength. No non-monotonic effects, of the type that feature so prominently in experimetal results \cite{raspaud},  are ever seen for any reasonable values of parameters ({\sl i.e.} the charges on the macroions, the length of the chain, the charges on the chain). Basing our conslusion on the anlysis presented above, we are inclined to believe that the polyelectrolyte bridging itself never leads to non-monotonic second virial coefficient. One nevertheless has to keep in mind, that our model calculation is based on  many constraints that are not entirely realistic. 

Our work represents an alternative formulation of the polyelectrolyte bridging interaction between two small macroions, on the two particle level. The finite size effects of the chain length make obviously a strong imprint on the bridging interaction. These effects have not been studied previously analytically and are missed by the more popular ground-state dominance mena-field approach. In this respect we believe our work adds an important feature to our understanding of the phenomenon of polymer mediated interactions.

{\bf Acknowledgment}

I would like to acknowledge very illuminating discussions with Eric Raspaud, Francoise Livolant, Amelie Lefoirestier, Stephanie Mangenot, E. Eisenriegler, Oleg Borisov, Theo Odijk, Luc Belloni and Per Lyngs Hansen. The present research was performed with the help of Chercheur Associe fellowship of C.N.R.S which is  gratefully acknowledged.

\eject

{\bf Figure captions}

Figure \ref{fig0}

A schematic representation of the model and the nature of the variational solutions. L.h.s.: the {\sl strong coupling} solution branch. The conformation of the chain is determined mostly by the interaction with the two macroions. It can be either in the {\sl weakly} (upper) or {\sl strongly paired} (lower drawing) configuration, depending on the separations between the macroions. The attractive bridging interaction is strongest in the {\sl strongly paired} configuration. R.h.s.: the {\sl weak coupling} solution branch. The conformation of the chain is determined mostly by the self-interactions of the chains. They are in the  {\sl weakly paired}  configuration for any separation between the macroions. On approach the interpenetration of the monomer clouds (the "coronas") leads to prevalently repulsive interactions. Note that the macroions are treated as point aprticles in the claculation and that their size is thus exagerated in this drawing.

\bigskip
Figure \ref{fig1} 

$\beta {\cal F}_0$ and $a$ in the strong coupling limit. The upper graph shows the dependence of $\beta {\cal F}_0$ for $M = 100$ for $N = 10 $ (losenge), $30$  (square), $100$  (circle) at $1 mM$  on the separation between the macroions $\vert {\bf r}_1 - {\bf r}_2\vert$. The lower graph shows the dependence of the size of the chain $a$ on the separation between the macroions for the same values of parameters.The length of the chain $N$ obviously determines the separation between macroions where the snapping of the chain between the strongly paired and weakly paired states occurs. In the strongly paired configuration we have a well developed regime of attractive bridging interactions leading to a an effective screened coulomb repulsion in the weakly paired regime. In all cases $\ell = 10 $. The bold line represents the pure Debye-H\" uckel interactions between the macroions. 

\bigskip
Figure \ref{fig2}

$\beta {\cal F}_0$ and $a$ in the strong coupling limit. The upper graph shows the dependence of $\beta {\cal F}_0$ for $M = 100$ for $N = 10, 30, 100$ at $60 mM$ on the separation between the macroions $\vert {\bf r}_1 - {\bf r}_2\vert$.  The lower graph shows the dependence of the size of the chain $a$ on the separation between the macroions for the same values of parameters. The bold line represents the pure Debye-H\" uckel interactions between the macroions.

\bigskip
Figure \ref{fig3}

$\beta {\cal F}_0$ and $a$ in the weak coupling limit for $M = 40$ and $N = 50, 20, 10$ at $1 mM$.  There exists only an {\sl weakly paired} state in this case and the effect of the external fields of the macroions is much less pronounced then in the strong coupling limit. The length of the chain $N$ determines primarily the steric repulsion effect due to the interpenetration of stiffened grafted chains, the "coronas" of both macroions, on close approach.  The inset shows the residual very weak coupling interaction at large separations between macroions with $N = 100$. The size of the chain tends to grow slightly on approach of the macroions because of the interpenetration of the "coronas" of both macroions. The bold line represents the pure Debye-H\" uckel interactions between the macroions.

\bigskip
Figure \ref{fig4}

$\beta {\cal F}_0$ and $a$ in the weak coupling limit  for $M = 40$ and $N = 50, 20, 10$ at $60 mM$. Clearly again the overall effect of the salt is to quench the effects of the macroion fields. There is weak residual bridging at very small separation now, note the scale of the interactions energy, that goes smoothly into a weak "corona" repulsion at larger separations. The details of both the residual bridging as well as "corona" repulsions depend on the values of $M$ and $N$. The bold line represents the pure Debye-H\" uckel interactions between the macroions.

\end{document}